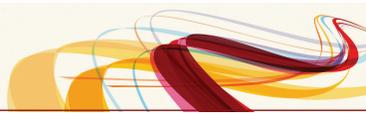

# Men Set Their Own Cites High: Gender and Self-citation across Fields and over Time

Molly M. King[1], Carl T. Bergstrom[2], Shelley J. Correll[1], Jennifer Jacquet[3], and Jevin D. West[2]

**Abstract**
How common is self-citation in scholarly publication, and does the practice vary by gender? Using novel methods and a data set of 1.5 million research papers in the scholarly database JSTOR published between 1779 and 2011, the authors find that nearly 10 percent of references are self-citations by a paper's authors. The findings also show that between 1779 and 2011, men cited their own papers 56 percent more than did women. In the last two decades of data, men self-cited 70 percent more than women. Women are also more than 10 percentage points more likely than men to not cite their own previous work at all. While these patterns could result from differences in the number of papers that men and women authors have published rather than gender-specific patterns of self-citation behavior, this gender gap in self-citation rates has remained stable over the last 50 years, despite increased representation of women in academia. The authors break down self-citation patterns by academic field and number of authors and comment on potential mechanisms behind these observations. These findings have important implications for scholarly visibility and cumulative advantage in academic careers.

Scholars of knowledge have long shown interest in citation analysis as a method for understanding academic professions (e.g., Adam 2002; de Solla Price 1965; MacRoberts and MacRoberts 1989; Newman 2001). One reason for this interest is that scholarly citations are an indicator of research performance, and research performance is an important part of the evaluation of faculty members when being considered for hire and promotion. Academia is unusual among professions in that performance may be relatively easily measured as research productivity or impact (van Arensbergen, van der Weijden, and van den Besselaar 2012). Evaluators also use various indices to measure whether a scholar's work is resonating with the discipline (Adam 2002), including citation counts.

When a scholar cites his or her own research ("self-citation"), this act may have a consequential impact on overall citations by both directly and indirectly increasing an author's citation counts. Not only does self-citation augment a paper's citation count by one, but on average, each additional self-citation yields an additional three citations from other scholars over a five-year period (Fowler and Aksnes 2007). Citation distributions are consistent with a preferential attachment model in which each citation received generates additional citations (Barabási and Albert 1999; Barabási et al. 2001; de Solla Price 1976; Peterson, Pressé, and Dill 2010; Redner 2005). Self-citations may be particularly important in this cumulative advantage process because they are often among the first citations to a paper (authors are the first to know about their work, after all).

The degree to which self-citation will create cumulative advantage across an individual scholar's career will vary across fields but can have a notable impact for some. In sociology, assistant professors commonly have 5 to 10 publications when they come up for tenure, while in biology, scientists commonly have 30 to 50 papers at tenure (and 15–30 papers when applying for assistant professor positions). Disciplinary differences in research organization, postdoctoral training, and publishing norms all affect the

[1]Stanford University, Stanford, CA, USA
[2]University of Washington, Seattle, WA, USA
[3]New York University, New York, NY, USA

**Corresponding Author:**
Molly M. King, Stanford University, Department of Sociology, 450 Serra Mall, Building 120, Room 160, Stanford, CA 94305-2047, USA.
Email: kingmo@stanford.edu

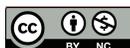

number of papers a scholar might have available to self-cite. Certainly, a scholar must write a reasonable number of papers for self-citation to have an effect early in his or her career; but in fields such as biology, self-citations can have an effect on career outcomes as early as the first job search, and certainly by the tenure stage. The cumulative record of citations (or lack thereof) from other scholars' citations that results from an author's self-citation patterns can seriously affect the appearance of scholarly influence.

Citation and self-citation rates appear to be precise measures of productivity and impact, but what if impact measures themselves contain nonmerit factors? In this article, we analyze 1.5 million academic papers from the JSTOR corpus to assess whether men academics cite their own papers more frequently than do women scholars.[1] If men are more likely to cite their own work, their papers will appear to have higher impact because of men's own (perhaps unconscious) efforts at self-promoting them, independent of any other qualities of the papers.[2] If women are less likely to self-cite, and if self-citation is a part of indices used to measure impact, then the tendency to self-cite less will be reflected in women's lowered impact measures (on average).

We further look at the gender patterns of self-citation over time. Two contradictory hypotheses are in tension here. Because the relative number of women in academia has grown over time (Hill, Corbett, and St. Rose 2010; National Center for Education Statistics 2013; National Science Foundation 2015c), we might expect gender gaps in self-citation to decrease. With more time in the profession, men have had more time to write papers and more time to cite the papers they have written. This implies that as women have been in the profession longer, the gender gap in self-citation should decrease. On the other hand, as academic jobs have become more competitive and the measures quantifying citations have become more important, scholars may feel more pressure to cite their own work as a way of boosting their own productivity ratings. If this pressure has caused men to be ever more likely to self-promote their work than women (Moss-Racusin, Phelan, and Rudman 2010), we might expect that gender gaps in self-citations would have increased over time.

In the next section, we review the literature on gender and academic research influence and, more specifically, the few smaller studies that have examined self-citation patterns. We then describe our data, which cover 1.5 million research papers dating back to 1779 (the date of the publication with the first self-citation). These are the largest and most comprehensive data ever used to examine gender and self-citation. We use a hierarchical classification algorithm to reveal the nested structure of fields and subfields. We assign gender to 2.8 million authors on the basis of first name, then calculate the rates of self-citation by gender within years and within fields. We also employ bootstrap methods to develop confidence intervals for our descriptive results. We find a substantial gender gap in self-citation in most fields. We finish by discussing several possible mechanisms underlying these observations and the important implications of these findings for academic institutions.

## Research Influence

The number of citations a paper receives (i.e., other articles that reference that particular work) is a common proxy for a publication's impact and influence. In the Web of Science, women in first- or sole-author positions receive fewer citations than men in the same positions (Larivière et al. 2013). In a study of articles published in the field of international relations between 1980 and 2006, papers in the same journal, published through the same peer-review process, are cited less often when written by women than when written by men (Maliniak, Powers, and Walter 2013). Natural science and engineering researchers who publish with a larger proportion of women coauthors are less cited than their colleagues who publish with more men coauthors, when targeting similar journals (Beaudry and Larivière 2016). Among sociology and linguistics faculty members at research institutions, men received more than twice as many cumulative citations to their articles as women, although much of this effect is explained by productivity differences (Leahey 2007). Another study of European molecular biologists found significant gender differences in the number of citations to a scholar's work only when the analysis was limited to first- and last-authored publications; there was no significant gender difference between total citation counts (Ledin et al. 2007).

Citation levels also depend on career stage and cohort. Among biochemists receiving their PhDs between 1950 and 1967, women's average number of citations per year is lower than men's at first, but by year 17, citation levels even out. In these same data, however, women have a higher average number of citations per paper: by career year 17, the average biochemist's paper is cited between 9 and 13 times if she is a woman or between 7 and 9 times if he is a man (Long 1992). In other words, this early cohort of senior women faculty members wrote fewer papers, but each paper was cited more than papers by men faculty members in equivalent positions. A study of 852 social scientists in the Netherlands found no gender differences in citation counts within the younger generation of researchers, compared with a significant gender gap in influence in the more senior cohort (van Arensbergen et al. 2012).

To date, there have been few studies of self-citation, and those that exist involved a limited number of disciplines and a relatively small number of papers, in part because publishers do not tend to provide free access to full-citation databases and because it is difficult to disambiguate author names. Only three studies on self-citation included analysis of gender. Research analyzing 12 journals in the field of international relations from 1986 to 2000 showed that men cite their own papers more than 1.5 times as often as do women (Maliniak et al. 2013). A study of papers in five archaeology publications also found that men tend to cite themselves slightly more often than women. However, this trend was not statistically significant, leading the author to conclude that there was no gender difference in self-citation (Hutson 2006). (The lack of significance in this study could have been due to the small sample size.) The third study examined 1,512 research-active authors who published in one of six leading ecology journals in 2011 and who began their publishing careers in 1994 or later. Of citations from the author to himself or herself, 8.5 percent of women's citations were self-citations, while 10.5 percent were men's self-citations, a significant 19 percent difference. The gender difference in self-citation among ecology researchers also significantly inflated some researchers' h-index, a metric used in career evaluation (Cameron, White, and Gray 2016).

---

[1]We intentionally use "men" and "women" rather than "male" and "female" to refer to scholars' gender (as reflected in their names) rather than biological sex.

[2]Self-promotion need not be a conscious strategy. Indeed, because self-promoting behaviors are discouraged for women but not for men, self-promoting behavior may be more common for men than women (Rudman et al. 2012). Even so, greater self-citation does increase citation count, thereby increasing the perceived quality or impact of a paper.

### *Inequality and Cumulative Advantage in Science Careers*

Given the importance of publication metrics in academic hiring, tenure and salary decisions, examining gender differences in citation patterns may shed light on persisting gender discrepancies in faculty hiring and promotion. Women remain underrepresented among tenured faculty members at U.S. universities, even though they have received the majority of bachelor's degrees for more than 30 years, and the number of women in postbaccalaureate programs has exceeded men nearly that long (National Center for Education Statistics 2013). In 2014, women earned 46 percent of all research doctorates, including 42 percent of science and engineering doctorate degrees (National Science Foundation 2015a). Even in a perfectly egalitarian hiring and promotion system, the lag in obtaining tenure means that it will take time to reach parity at tenured ranks. In the social sciences, in which women have earned PhDs at a higher rate than men for two decades (National Science Foundation 2015a, 2015b), we still see women underrepresented in faculty positions (National Science Foundation 2015c). Furthermore, women are underrepresented in senior ranks of faculty, even after controlling for factors such as experience (reviewed in Bentley and Adamson 2004). Among doctoral scientists and engineers at 4-year institutions in 2013, 28 percent of tenured faculty members were women, compared with 42 percent on the tenure track and 46 percent who were not in tenure-track positions (National Science Foundation 2013). Women are also underrepresented as faculty members at the most elite universities (National Science Foundation 2015c), even after controlling for research productivity and department factors (Weisshaar forthcoming).

At institutions offering tenure in 2011 and 2012, 54 percent of men but only 41 percent of women full-time instructional faculty members had tenure (National Center for Education Statistics 2013). Controlling for numbers of papers authored as well as other institutional factors, women assistant professors are still less likely their men counterparts to receive tenure (Weisshaar forthcoming). These status differences translate to real-world economic outcomes. Most studies show that women faculty members earn less than men faculty members (reviewed in Bentley and Adamson 2004). In the 2012–2013 academic year, men faculty members earned about 22 percent more than women faculty members at degree-granting two- and four-year institutions (average salary $84,000 vs. $69,100; National Center for Education Statistics 2013).

As Rossiter (1993) documented, women's academic contributions to science have been undervalued historically. She referred to the process by which women's scientific contributions are downplayed or ignored relative to men's as the "Matilda effect."[3] This phrase contrasts with the well-known "Matthew effect," which refers to the psychosocial process of cumulative advantage, by which eminent scientists receive credit disproportionately to their contributions (Merton 1968, 1988). There is then a "continuing interplay between the status system, based on honor and esteem, and the class system . . . which locates scientists in differing positions within the opportunity structure of science," providing eminent scientists with further advantages in the quest to contribute (Merton 1968:57). Recognition is a primary source of barter and reward in scientific careers, underscoring the importance of understanding citation patterns as part of the Matthew and Matilda effects.

New citations are statistically more likely to accrue to those papers that are already the most cited (Barabási et al. 2001; de Solla Price 1976). Again, self-citations aid this process by encouraging future citations from other scholars (Fowler and Aksnes 2007). Because self-citation represents a nontrivial component of all academic citations, as we show below, it is important to understand if there are systematic gender patterns in self-citation across a broad range of fields. Such patterns may mechanically disadvantage one gender and contribute to cumulative advantage of rewards that undergirds a successful career in academia.

## Methods

### *Self-citations: An Author-to-author Approach*

Disambiguating authors (i.e., determining when multiple papers are written by the same individual and when they are written by different individuals with the same name) is one of the major challenges in bibliometric analysis (Smalheiser and Torvik 2009). The JSTOR data set is not disambiguated.[4] To tally self-citations without author disambiguation, we assume that any citation to an author with the same name is a self-citation. There are a vast number of possible combinations of first and last names and a relatively small number of papers that will be cited as references on a paper in comparison. Given this, we think it safe to assume that all but an inconsequential number of citations from an author John Williams to a published paper by a John Williams will be self-citations in their intended sense—they were written by the same individual, not just by two individuals who just happen to have the same name.

---

[3]Evidence of such gender differences in evaluations of scientific contributions is also perceived in gender-differentiated ways. Results from three different experiments, using samples of both public and scientific communities, showed that men evaluate evidence of gender bias in science as less meritorious than do women (Handley et al. 2015).

A bigger problem is that because we cannot track individual authors over time, we cannot control for differences in career stage or individual productivity. For example, men authors may, on average, have more papers they can self-cite than do women authors. This could, in principle, generate a gap in self-citation rates even if men and women with the same number of published papers self-cite at identical rates.

When tallying self-citations, we consider all author-to-author citations, whereby a paper with four authors citing a paper with three authors counts as 12 author-to-author citations, one for each combination. For example, a paper written by four authors Pooja Joshi, Colin Edwards, Armand Erickson, and John Williams (2010) cites a paper written by three authors Rita Martin, Colin Edwards, and Sarah White (2008). Colin Edwards (but no one else) is an author on both papers. This citation represents 12 author-to-author pairs (Joshi to Martin, Joshi to Edwards, Joshi to White, Edwards to Martin, etc.) of which one, Colin Edwards to Colin Edwards, is a self-citation. Thus 1/12 of the author-to-author citations here is considered a self-citation. The fraction of author-to-author self-citations will always be smaller than or equal to the fraction of citations that can be considered as paper-level self-citations. Our example illustrates this plainly. At the paper level, the sole citation listed, from Joshi, Edwards, Erickson, and Williams (2010) to Martin, Edwards, and White (2008), is considered a self-citation because Colin Edwards is on both papers. Therefore, although 1/12 of the author-to-author citations are self-citations, 100 percent of the paper-level citations are self-citations.[5]

An *authorship* is a unique author-paper pair. In the example above, an authorship would be Pooja Joshi's 2010 paper. If Joshi wrote more than one paper in 2010, each would be a separate authorship. If there are several authors on a paper, there can be multiple authorships per paper. In the above example, the 2010 paper by Joshi, Edwards, Erickson, and Williams involves four separate authorships.

We define the *self-citation rate* as the mean self-citations per authorship.[6] Let $a$ be the number of authorships and $s$ be the number of self-citations for a given group (across a year, gender, etc.). So across a group of papers, the mean self-citation rate will be the total number of self-citations out of the total number of authorships. Let $a_w$ and $a_m$ be the number of women's and men's authorships, respectively. Let $s_w$ and $s_m$ be the number of women's and men's self-citations, respectively. At what rate $k$ do men self-cite? We calculate the relative rate $k$ of men's self-citation to women's self-citation by solving the following expression for $k$:

$$\frac{s_m}{a_m} = k \frac{s_w}{a_w}.$$

In calculating the self-citation rate, we match using full first and last names of authors and cited references, disregarding any middle initials. The date of a self-citation is taken to be the citing year, rather than the cited year.

---

[4] Disambiguation would highlight ties between papers by identifying when the same name belongs to the same individual across different authorship instances. We could have fully disambiguated the authors on a very small number of papers by hand. However, having only a small sample would rule out assessing self-citation trends across many fields and across such an unprecedentedly large amount of time. We felt that we could best contribute by exploiting the longitudinal and cross-disciplinary nature of these data.

[5] What if an author now has a hyphenated name due to marriage (e.g. Smith-Johnson) but references an article written under his or her previous nonhyphenated name (e.g., Smith)? Hyphenated names due to marriage are not of significant concern in our network data set: there are only 51,270 authorships with hyphens (1.8 percent of the total), with only a fraction of these likely due to the author's marriage. Similarly, we cannot evaluate the impact of complete last name changes due to marriage in our data, whereby a self-citation would go unrecognized because a person's last name has changed. This would likely result in a slight downward bias of our estimate of women's self-citation rates. We do not believe that this effect has a major impact on our results, however. In the past two decades, we find that men cite themselves 70 percent more, compared with 57 percent more across the more than two centuries of our full data set. With women academics taking the surnames of their spouses less frequently in recent years, we would instead predict a substantial reduction in this percentage if married names were having a major influence on our findings.

[6] The self-citation rate as defined here measures the fraction of the outgoing citations an author makes that go to his or her other papers. It would be extremely interesting to look at the fraction of incoming citations an author receives that come from his or her own papers, but without the ability to disambiguate authors, we are unable to consider this metric in the present paper.

*King et al.* 5**Table 1.** Network and Analytic Data Set Sizes on the Basis of Various Descriptors of Papers, Citations, and Authorships.

| Data Set | Description | | Value |
|---|---|---|---:|
| Network data set | Papers | | |
| | | Papers (including both citing and cited) | 1,787,351 |
| | | Unique citing papers that cite other JSTOR papers | 1,388,431 |
| | | Unique citing papers that self-cite | 411,403 |
| | Citations | | |
| | | Paper-to-paper citations | 8,227,537 |
| | | Paper-to-paper citations that are self-citations | 774,113 |
| | | Author-to-author citations | 39,402,992 |
| | | Unique citing-cited pairs of author-to-author citations | 6,268,789 |
| | Authorships | | |
| | | Total authorships (paper-author pairs) | 3,578,138 |
| Analytic data set | Papers | | |
| | | Papers with extractable author names | 1,450,605 |
| | | Unique citing papers with author names that cite other JSTOR papers | 1,092,376 |
| | Authorships | | |
| | | Authorships (paper-author pairs) with author names | 2,787,833 |
| | | Men authorships, 1779–2011 | 1,595,721 |
| | | Women authorships, 1779–2011 | 448,386 |
| | | Men authorships, 1950–2011 | 1,501,312 |
| | | Women authorships, 1950–2011 | 435,396 |

*Note:* All data derive from the JSTOR database from 1779 to 2011 unless otherwise noted.

## The JSTOR "Network Data Set"

JSTOR is a not-for-profit digital collection of scholarly documents ranging in time from the mid-sixteenth century to the present day. The JSTOR collection includes over 8 million individual documents and over 4 million research articles, of which 1.8 million are linked by citation to other articles in the collection. We focus on these documents, which we call the JSTOR "network data set," because they are amenable to citation network analysis. We obtained the citation data and full-text publication data from JSTOR. We used these data to determine author gender and to calculate self-citation rates. We then constructed a citation network, which we used to determine disciplines (e.g., ecology).

We include only papers published in or after 1779, the year of the first self-citation in the JSTOR corpus, making our analytical data set approximately 1.5 million papers. Our analyses are based on the 1779–2011 data set, although sometimes we report only the years after 1950 or 1970, when sample sizes from earlier periods would be too small to draw any meaningful conclusions, which is explicitly noted.

There are 3.6 million total authorships in the network data set and more than 39 million author-to-author citations. There are 6.2 million unique citing-cited pairs of author-author citations in the network data set. Therefore, among the 39 million author-to-author citations, many pairs occur repeatedly (as might be expected when a paper cites multiple papers by the same author). The network data set also includes 8.2 million paper-to-paper citations. Of these, more than three quarters of a million paper-to-paper citations are self-citations. Further descriptive detail for this dataset can be seen in Table 1.[7]

## Mapping the Hierarchical Structure of Scholarly Research

To analyze self-citation differences across academic fields, we used hierarchical classification to reveal the structure of fields, subfields, and ever finer partitions down to the level of individual research topics. A prior analysis (West, Jacquet, et al. 2013) used the hierarchical map equation (Rosvall and Bergstrom 2011) to create a nested hierarchy of all papers in this network data set on the basis of citation relations among the papers. The hierarchical map equation algorithm determined the boundaries between groups at each level of the hierarchy. We manually assigned names to the field, subfield, and research topic groups that the algorithm revealed by examining the 50 most important papers in each of the groups (based on the number of citations to each paper).

The hierarchical map equation leverages the duality between compressing data and finding patterns in those data. When one compresses a night view image of a country, the major highways and cities are highlighted. We compress citation networks in a similar way. But instead of roads and cars, our map shows citation trails (when a paper cites a reference paper) and the ideas transmitted along those citation trails. After releasing a random walker on the network, the algorithm tries to minimize the description length of the random-walk process. In areas of the network where the random walker spends extra time moving back and forth within the same group of papers, the algorithm assigns an "area code." These area codes that the random walker reveals are fields of science. These methods have been vetted in the network science literature and consistently outperform other community detection algorithms (e.g., Aldecoa and Marín 2013; Lancichinetti and Fortunato 2009; Moody and White 2003; Šubelj, van Eck, and Waltman 2016). The open-source code for running the hierarchical map equation is called InfoMap and can be found at http://mapequation.org/code.html. In this article, we use the article-level eigenfactor (West, Rosvall, and Bergstrom 2016) as the underlying random-walk process that the hierarchical map equation compresses. This is a modified version of PageRank that is customized for article-level citation networks and works well for ranking nodes and revealing hierarchical structure (Wesley-Smith, Bergstrom, and West 2016). Again, we used this community detection method of hierarchical classification to identify the academic field of each paper.

---

[7]The number of references recorded per article on average has increased from about three in the 1950s to more than nine in the 2000s.

### Determining the Gender of Authors: The "Analytic Data Set"

We extract the first names of authors from 1.5 million papers in the JSTOR network data set. To assign gender to first name, we use the methods of West, Jensen, et al. (2013), which relied on U.S. Social Security Administration records (available at http://www.ssa.gov/oact/babynames/) to provide information about first names and corresponding gender.[8] We assign gender to authors' names that appear in the top 1,000 most popular names in any year from 1879 to 2012.[9] We assume that we can confidently assign gender to author if the author's first name has the same gender at least 95 percent of the time in the Social Security database.

Authors with first names that are associated with both genders, such as Jody and Shannon, were dropped from the analysis, as were authors listed only by their first initial. Disregarding authors with only first initials may exclude female authors disproportionately, particularly in early eras when women may have been more likely than men to publish with initials to avoid potential discrimination. Because in any given era, gender-ambiguous names are more likely to be women (Lieberson, Dumais, and Baumann 2000), this may slightly downwardly bias our appropriate assignments of women. Similarly, we were unable to classify names that were not in the top 1,000 Social Security Administration records for any year from 1879 to 2012. As a result, authors of some nationalities may be underrepresented in our data set. In a few rare cases, national differences may cause misleading assignments for non-U.S. authors (e.g., Andrea is typically a woman's name in the United States but a man's name in Italy).

As discussed above, an instance of authorship consists of a person and a paper for which the person is designated as a sole author or coauthor. Of the 3.6 million authorships in the JSTOR network data set, we were able to extract a full first name for 2.8 million authorships (77 percent). We were able to confidently assign gender to 73.3 percent of these authorships with full first names, including 1.6 million men and nearly 0.5 million women. The remaining authorships involve names not in the Social Security lists (24.3 percent) or names associated with both genders (2.4 percent).[10] The final analytic data set includes all papers for which we know the gender of one or more authors. The values for these different data types, and others, can be seen in Table 1.

### Bootstrapping Standard Errors

We use bootstrap methods to estimate confidence intervals for our self-citation rates and ratios. We resample papers, with replacement, from the appropriate sample set (year, field, or year and field) and calculate our statistics of interest on the total of all authorships within all resampled papers. To estimate confidence intervals for the ratios, we resample at the level of individual papers, calculate self-citation rates (in each bootstrap sample), and then take the ratio of these rates in our network data set. Two different authors used Stata/IC 13.1 for OS X and Mathematica 10.1 for OS X to independently confirm bootstrap results.

---

[8]We are therefore restricted to following the Social Security Administration data, which acknowledge only two genders.

[9]The JSTOR data set includes papers to 1750, but we only have names from 1879 from the Social Security Administration records. However, this will likely not change the results. First, there are relatively few papers before 1850, so this proportion of the data has little effect on the overall results. Second, most names prior to 1879 exist in the data set from 1879 to 2012. The minor changes in frequencies will have little effect on the results prior to 1879.

[10]The fraction of authorships with unknown gender is typical for a data set of our size (Larivière et al. 2013). Efforts to increase coverage would come with a loss of accuracy. A study comparing our approach of name matching on the basis of Social Security records with others found that including data sets from other countries, manual coding of names, or a unisex category might produce more biased results (Wais 2016). Specifically, these techniques "will not necessarily increase the proportion of items with predicted gender and can also contribute to the bias of gender proportion estimates" (p. 35). Among the approaches studied, Wais (2016) found a trade-off between the goals to predict gender for as many people as possible and maximizing prediction accuracy. Of the three approaches studied, ours fell in the middle, actually having a lower error rate than that which included manual coding of gender.

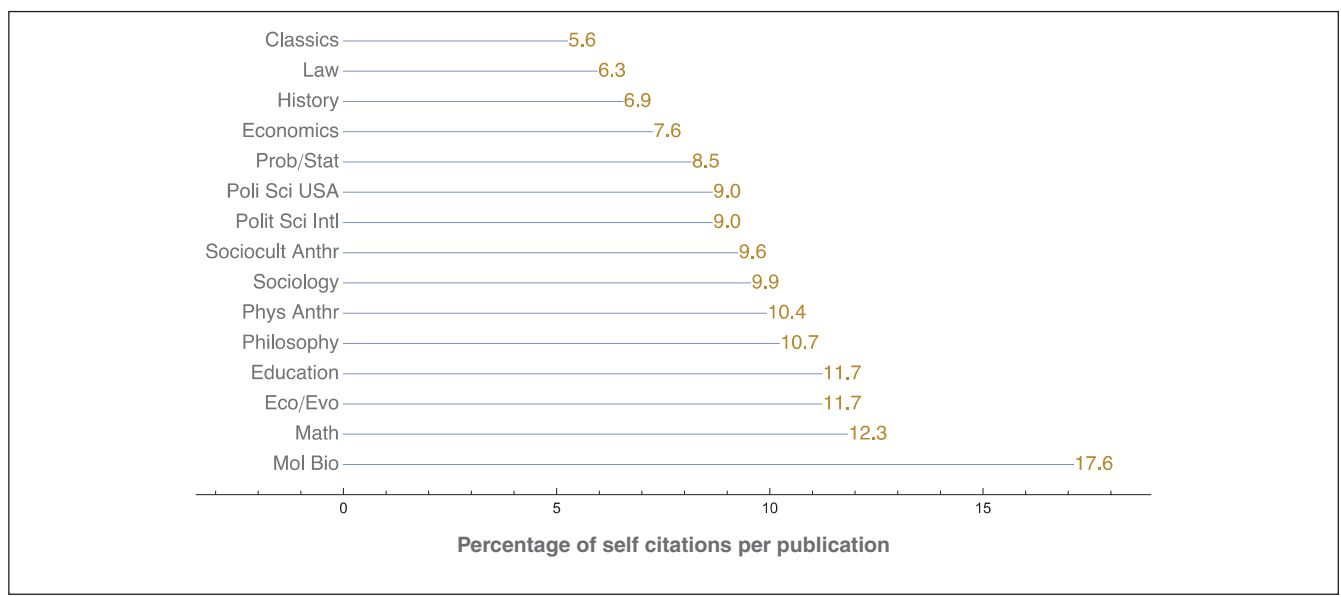

**Figure 1.** Mean percentage of self-citations per publication by field in JSTOR, 1779–2011. Shown here is the percentage of a paper's references that cite papers written by one or more of that paper's authors, averaged across each major field. A value of 10 means that 10 percent of a paper's citations are references to papers previously written by one or more of the paper's authors.

Our original sample set contains $m$ men and $w$ women authorships. We draw $n$ bootstrap samples, each with $m$ men and $w$ women authorships selected with replacement from the original data. For each bootstrap sample $i$, we compute the men's self-citation rate $\hat{\bar{x}}_i$ and the women's self-citation rate $\hat{\bar{y}}_i$. For each $\hat{\bar{x}}_i$ and $\hat{\bar{y}}_i$ we compute the bootstrap ratio of men's self-citations to women's self-citations:

$$\hat{k}_i = \frac{\hat{\bar{x}}_i}{\hat{\bar{y}}_i} \text{ for all } i = 1, \ldots, n.$$

We then order all $\hat{k}_i$ such that $\hat{k}_i \leq \hat{k}_{i+1}$ and find the value of $\hat{k}_i$ at the 2.5th percentile and the 97.5th percentile of $n$. These values are the lower and upper bounds of the 95 percent bootstrap confidence interval.[11]

### Results

#### How Common Is Self-citation?

To provide more context for the importance of self-citation, we wanted to know what proportion of citations in an article are self-citations, on average. This helps address the relative importance of gender disparities without disambiguating author names. Within all papers in the JSTOR corpus, 774,113 paper-to-paper references were self-citations, meaning that 9.4 percent of 8.2 million references were self-citations. Put another way, across all fields and years, about 1 in 10 references is a self-citation.[12]

Figure 1 presents these results broken down by major academic field. Molecular biology has the highest self-citation rate per reference, while classical studies has the lowest.

The paper with the most self-citations by its authors is a report in *Science* titled "A Comparison of Whole-genome Shotgun-derived Mouse Chromosome 16 and the Human Genome." In no sense is this paper an example of excessive self-citation; the paper references only four previous papers written by any of the paper's 175 authors. But because three of the cited papers each have many authors from the citing paper, the authorship-to-authorship links add up to 220 self-citations. Another example is a paper in the *American Economic Review* titled "Information and the Change in the Paradigm in Economics." This is single-authored paper with 70 self-citations out of 130 references. This is likewise not a case of excessive self-citation, because the paper is an adaption of Joseph Stiglitz's Nobel Prize lecture, the whole point of which is to trace the arc of his career.

However, these two papers illustrate alternative paths to the same end: at one extreme, papers with many authors citing even a few papers with many of the same authors; at another extreme, sole-authored papers citing many previous papers. These different effects may be differentially likely in different fields. Our analysis did not suggest any notable

---

[11]For example, if $n = 10{,}000$, after sorting the values in ascending order, the 2.5th and 97.5th percentiles of the distribution fall at positions 250 and 9,751.

[12]Within only those papers that included self-citations, there was a total of 3,754,942 references. Among only these papers that cite earlier papers written by their same authors, then, approximately 21 percent of included references are self-citations.

relationship between the number of references cited by a paper and the number of self-citations.

Self-citation can be an influential force in raising an academic's citation count. For a powerful example, consider one prominent scholar—listed by Thomson Reuters as one of its Highly Cited Researchers—with nearly 7,000 Web of Science citations. Of these, more than 1,500 are self-citations.[13] On average each of this author's more than 290 papers cites slightly more than 5 of his previous papers. As a result, this scholar receives nearly 22 percent of his citations from himself, even ignoring the additional citations from others that are generated by preferential attachment processes (Fowler and Aksnes 2007). Although this is obviously an extreme case, and it is not our aim to criticize the practice of self-citation,[14] we do want to emphasize how common self-citation is, along with the profound effect it can have on an academic's citation count.

## Self-citation Patterns by Gender

Between 1779 and 2011, there are 1,595,721 men authorships and 448,386 women authorships in our analytic data set. Men represent 78.1 percent and women 21.9 percent of authorships for which we could identify the gender, dating back to 1779. Dating back to 1950, there are 1,501,312 men authorships and 435,396 women authorships. Since 1950, men represent 77.5 percent of the authorships for which we know the gender, and women make up the remaining 22.5 percent. There were 743,319 authorships for which we could not identify gender. Moving the start of the window from 1779 to 1950, then, we see a change in the authorship gender gap by less than 1 percentage point. The change is so slight because JSTOR contains comparably few documents dating to before 1950.

Because papers often have more than just one author, author-to-author citations outnumber paper-to-paper citations. In the analytic data set, there are 1,017,362 author-to-author self-citations. Of these, there are 678,768 self-citations by men, 121,923 self-citations by women, and 216,671 self-citations by authors of unknown gender. This means that of the self-citations for which we know the author's gender, men are responsible for 84.8 percent of the self-citations, while women are responsible for 15.2 percent of the self-citations.

Standardizing women's self-citation rate to 1.0, we solve for the ratio of men's self-citations relative to women's for 1779 to 2011:

$$\frac{\% \text{ men authorships} \times \text{men's selfcite rate } k}{\% \text{ women authorships} \times \text{women's selfcite rate}} = \frac{\% \text{ men's selfcites}}{\% \text{ women's selfcites}},$$

$$\text{or } \frac{78.1 \times k}{21.9 \times 1} = \frac{84.8}{15.2}$$

Solving for $k$, we find a ratio of 1.56, meaning that the average man self-cites 56 percent more often than does the average woman. This is remarkably consistent with the results reported by Maliniak et al. (2013), who analyzed 3,000 articles from the field of international relations and reported that men authors self-cite 60 percent more often than women authors. (Using the JSTOR network data set, we find that men in the field of (U.S.) domestic political science self-cite their own work 58 percent more often than women, and, in international political science, 68 percent more often.)

Next we visualize the total number and fraction of self-citations by author gender. We look at absolute numbers rather than the percentage of a paper's citations that are self-citations, because there are many papers with one citation that is a self-citation; visualizing the percentage of citations that are self-citations results in long tails and does less to further our understanding.

In how many papers do men and women authors cite themselves $n$ times? Figure 2 shows the log frequency of self-citation counts by gender for each number of self-citations. Men have higher counts in all categories of numbers of self-citations, including papers with no self-citations, which is not surprising since there are more instances of men authorship in the network data set.

Figure 3 helps us explore these numbers further using relative proportions. It shows members of self-citations grouped by proportions of men's and women's authorships. We show the proportion of men with a certain number of self-citations on the x-axis and the corresponding proportion of women on the y-axis. If men and woman behaved similarly in their approaches to self-citation, the corners of the boxes would trace the x-y diagonal. Instead, wherever there is a difference in the proportion of men and women citing themselves a certain number of times, the corners of the boxes deviate from the diagonal.

Figure 3 shows us that relative to men's authorships, women's authorships are more likely to feature zero self-citations. Women cite themselves one or more times in their papers less often than men do. In other words, compared with men, women are overrepresented in the zero self-citations category and underrepresented in terms of citing their papers at all. For example, if in a paper you never cite another paper of your own, you are among the majority of men (68.6 percent) and women (78.8 percent) who do not cite themselves.

In fact, we can see from Figure 3 that whenever a box is wider than it is tall, there is a greater proportion of men authorships in that category of self-citations. If you have one self-citation, you are in the 68th to 88th percentile range for men (representing 20 percent of men's authorships) but the 78th to 93rd percentile for women (representing only 15 percent of women's authorships). With four self-citations in a single paper, a woman is in the 99th percentile, while a man is in the 98th.

---

[13]Note that because names are not disambiguated in JSTOR, we cannot check for similar extremes in our own analyses.

[14]For example, the present article will provide the authors with 1, 5, 3, 1, and 4 self-citations, respectively, by authorship order; and although we believe that none of the self-citations herein are extraneous, we note that the men authors of this article cite themselves at nearly three times the average rate of the women authors.

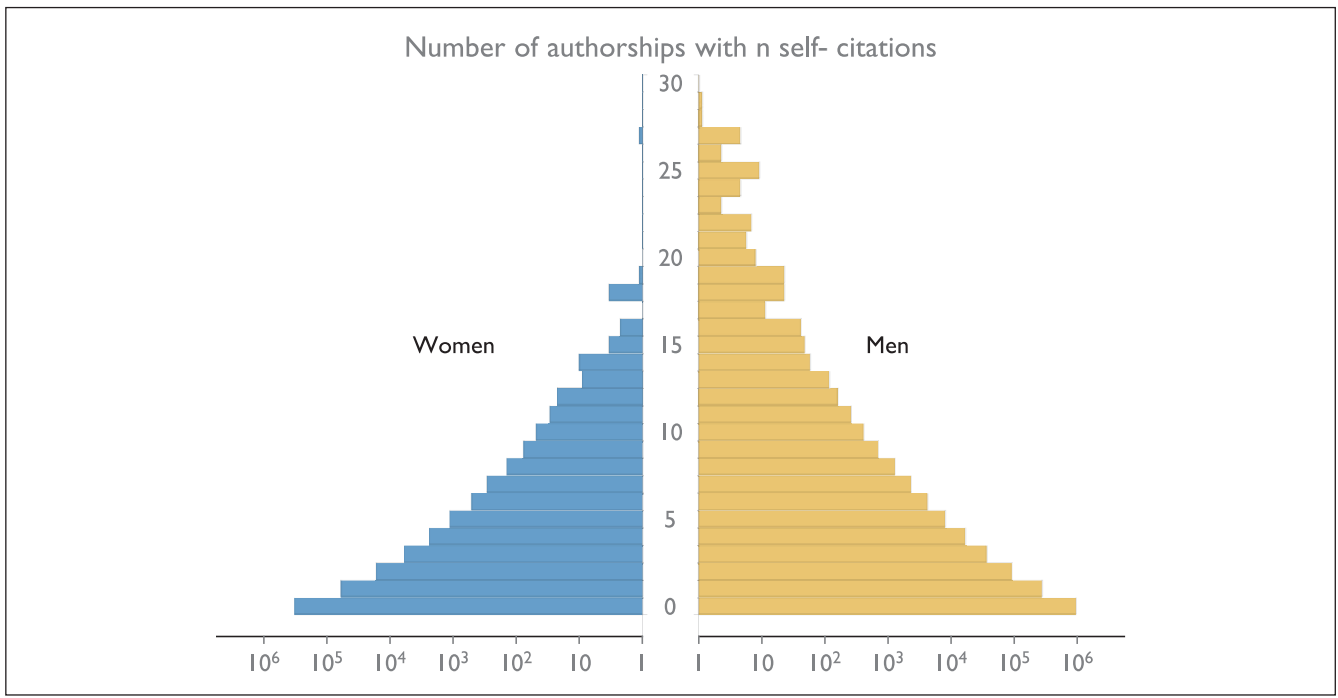

**Figure 2.** Number of authorship instances in which specified number of self-citations (per authorship) occurs, by gender in JSTOR, 1779–2011. Each bar's length (along the horizontal axis) is equal to the log number of observations of *n* self-citations (on the vertical axis).

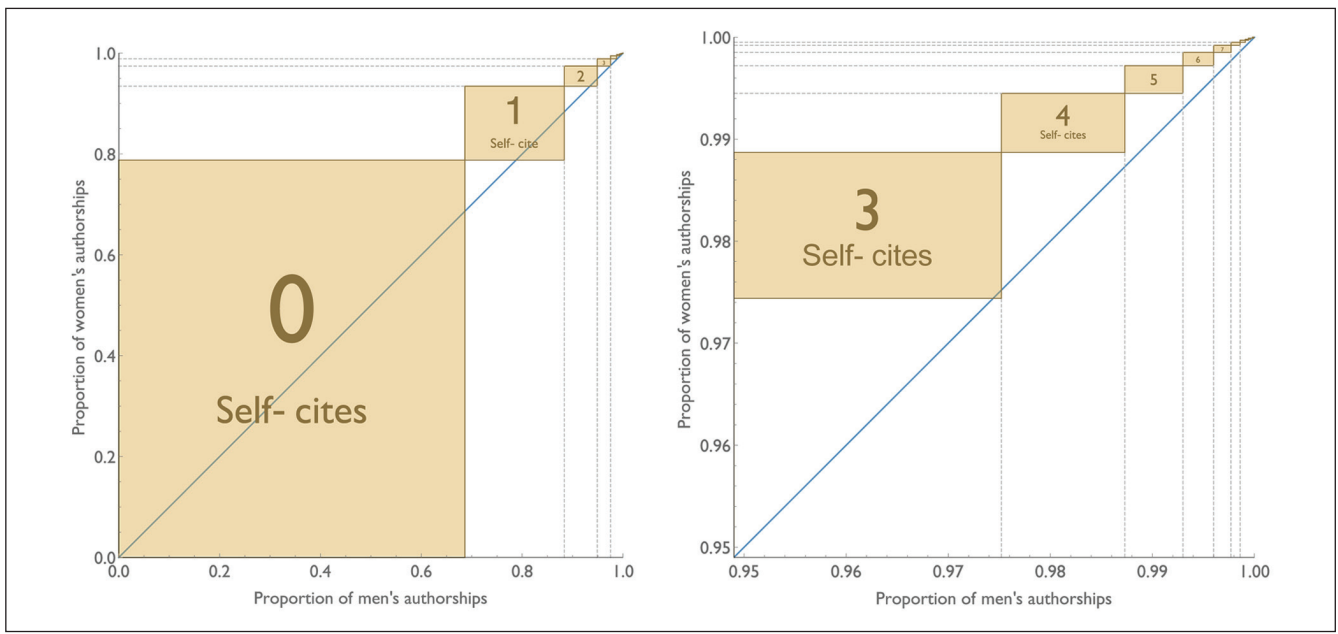

**Figure 3.** Proportion of authorship instances in which specified number of self-citations occurs, by gender in JSTOR, 1779–2011. It was produced by converting the count of authorship instances in which specified number of self-citations (per authorship) occurs by gender (Figure 2) into proportions. The first half of the figure shows the whole range of possible numbers of self-citations, while the second half zooms in on the area representing three self-citations and above. The right edge of each box indicates the proportion of men who cite themselves that number of times, while the upper edge of each box indicates the proportion of women who cite themselves that number of times. The diagonal line represents the point of gender parity, which would bisect the boxes through their corners if the genders behaved identically in patterns of self-citation.

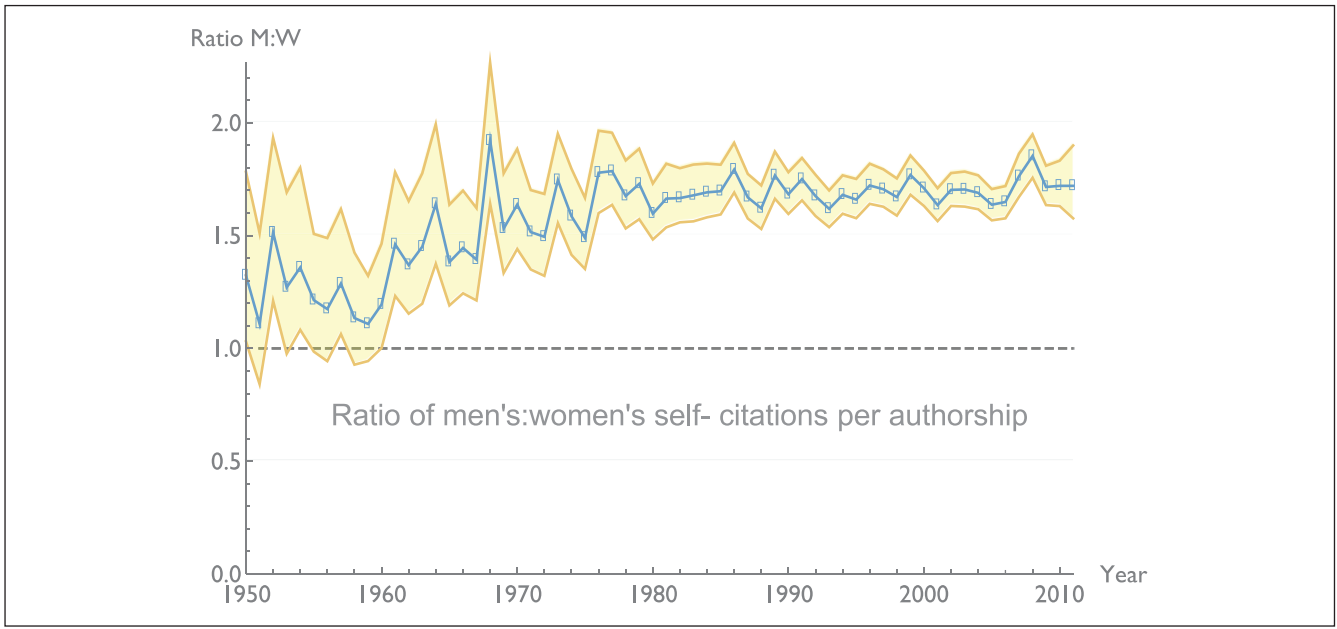

**Figure 4.** Men cite themselves more than women do. Shown here, the ratio of men's self-citations per authorship relative to women's self-citations per authorship, for JSTOR articles over the period from 1950 to 2011. If men and women cited themselves at equal rates, the ratio shown would be 1.0. A value of 1.5 means that men cite themselves 50 percent more than women in papers published during that year. Shaded intervals represent 95 percent bootstrap confidence limits.

Understanding these distributions is important because they help us see that the gendered nature of self-citation averages is not a result of highly skewed tails representing aberrant behavior. It is the product of the daily activity of the vast majority of academics, those who cite themselves in their papers fewer than five times.

### Self-citation Rates over Time

The very first self-citation in our data set was in 1779 to a paper dated 1773. Edward King, in his paper "Account of a Petrefaction Found on the Coast of East Lothian" (King 1779) cites his own previous "A Letter to Mathew Maty, M.D. Sec. R S.; Containing Some Observations on a Singular Sparry Incrustation Found in Somersetshire" (King 1773).

Figure 4 shows the self-citation ratio for each year. In the 1950s, the relative rate[15] of men's self-citations relative to women's self-citations was 1.23. However, during the 1950s, the bootstrapped 95 percent confidence intervals of the annual ratios overlap with an equality ratio of 1.0, indicating that we cannot reject the null hypothesis of gender equality in self-citation rate during this decade. However, beginning in the 1960s, the ratio of men's to women's self-citations per authorship remains steadily significantly above 1.0. In the 2000s, the relative rate was 1.71. There is no evidence that that the gender gap is decreasing over time.

Because the ratio is composed of the relative rates of men's and women's self-citations, we wondered what the patterns underlying this trend might be: Do both men and women self-cite at increasing rates? Or are the rates for each gender relatively steady over time? To investigate this, we plotted men's and women's self-citation rates separately over time (Figure 5).

Beginning in the 1960s, men had a consistently higher rate of self-citation than women did, across all fields. Note that the sharp drop after 2006 is likely due to the blackout

---

[15]The relative rate is calculated by first summing the total number of self-citations by men (or women) across the decade, then dividing this by the sum of the total number of men (or women) authorships across the decade:

$$\text{Men's 1950s rate} = \sum_{1950}^{1959} S_y^M \bigg/ \sum_{1950}^{1959} A_y^M$$

where $S_y^M$ is the number of self-citations by men in year $y$ and $A_y^M$ is the number of men authorships in year $y$. The men's rate for the decade is then divided by the women's rate for the decade to give the relative rate. This is important because the sample sizes differ in each year and because the relative contribution of each year may differ for men and women. We compute the average rates across the decade for each gender and only then take their ratio.

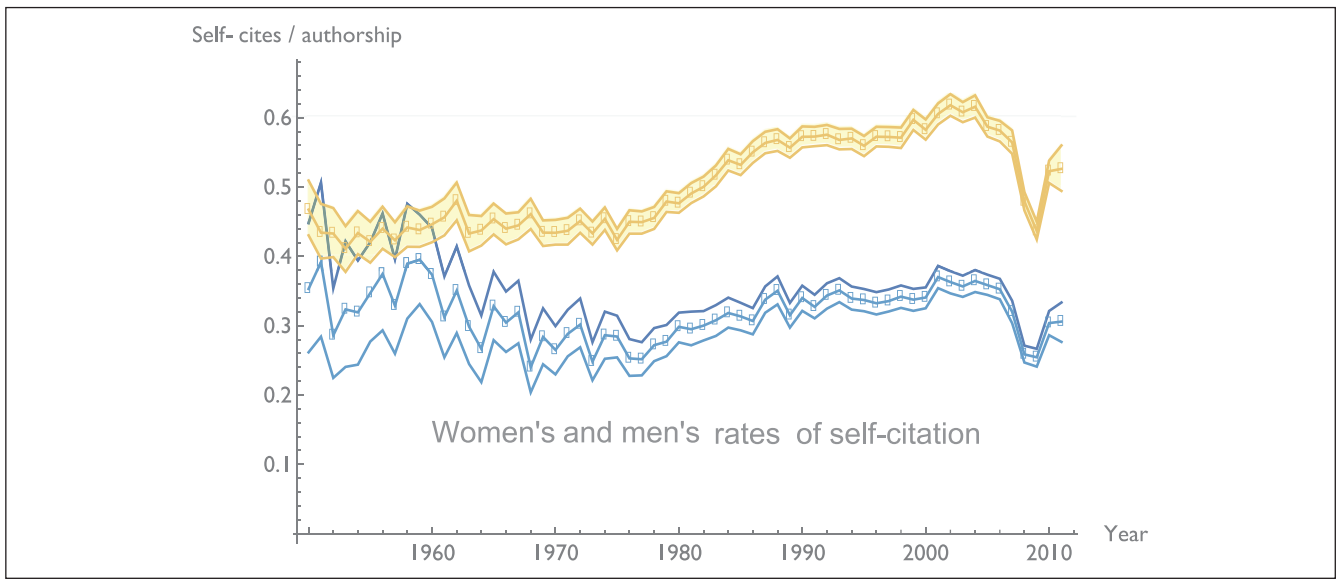

**Figure 5.** Men's rate of self-citation has been higher than women's since the 1960s. Shown here, the mean number of men's self-citations per authorship (yellow line) and women's self-citations per authorship (blue line), for JSTOR articles over the period from 1950 to 2011. Shaded intervals represent 95 percent bootstrap confidence limits.

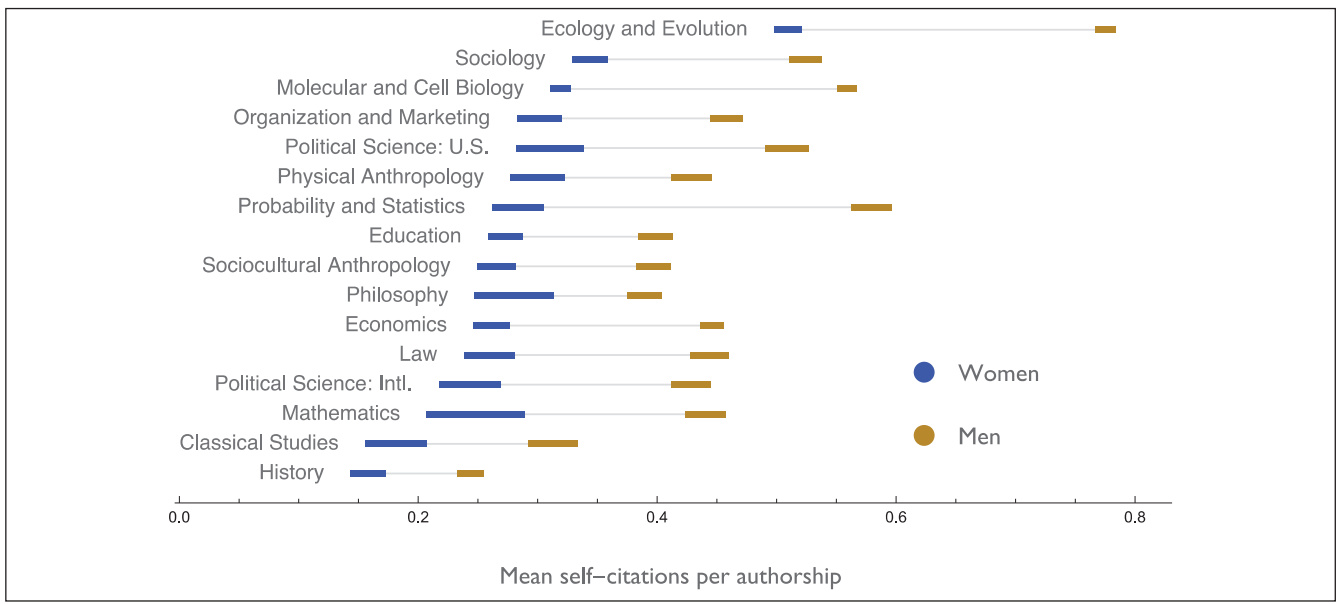

**Figure 6.** Mean number of men's self-citations and mean number of women's self-citations per authorship across major fields, based on author-to-author self-citations, in JSTOR, 1779–2011. Orange numbers represent men's average number of self-citations per authorship in that field, and blue numbers represent women's average number of self-citations per authorship. Dark-colored bars represent 95 percent bootstrap confidence intervals for each gender.

window for certain fields (some papers do not appear on JSTOR until five years after publication), combined with differences in self-citation rates across fields.

### Self-citation Rates by Field

Although the average ratio shows that men cite their own papers more than women, self-citation behavior varies widely across fields and subfields. Figure 6 shows men's and women's self-citation rates by major academic field. Each and every field in the plot reveals a large and significant difference between women's and men's self-citation rates.

Previous research found that women's disadvantage in garnering citations decreased as women made up an increasingly large proportion of the field of economics (Ferber and Brün 2011). We wondered whether the gender composition of a field's authorships might correlate with the rate of self-citation. The fields with the lowest women's self-citation rates per authorship (and their corresponding proportions of women authorships in each field from 1779 to 2011) are history (22.5 percent) and classical studies (22.3 percent). The fields with the highest women's self-citation rates per authorship are ecology and evolution (19.4 percent), sociology (32.9 percent), and molecular and cell biology (26.8 percent). Under a linear model there is no significant relationship between women's (or men's) self-citation rate per authorship and the proportion of authorships that are women in a field. It is possible that what might matter more than a continuous level is some threshold level at which women are no longer considered tokens in the workplace (Cain and Leahey 2014; Kanter 1993). Although our measure is a continuous one, our analysis shows little evidence of a threshold effect here, either (see Appendix B).

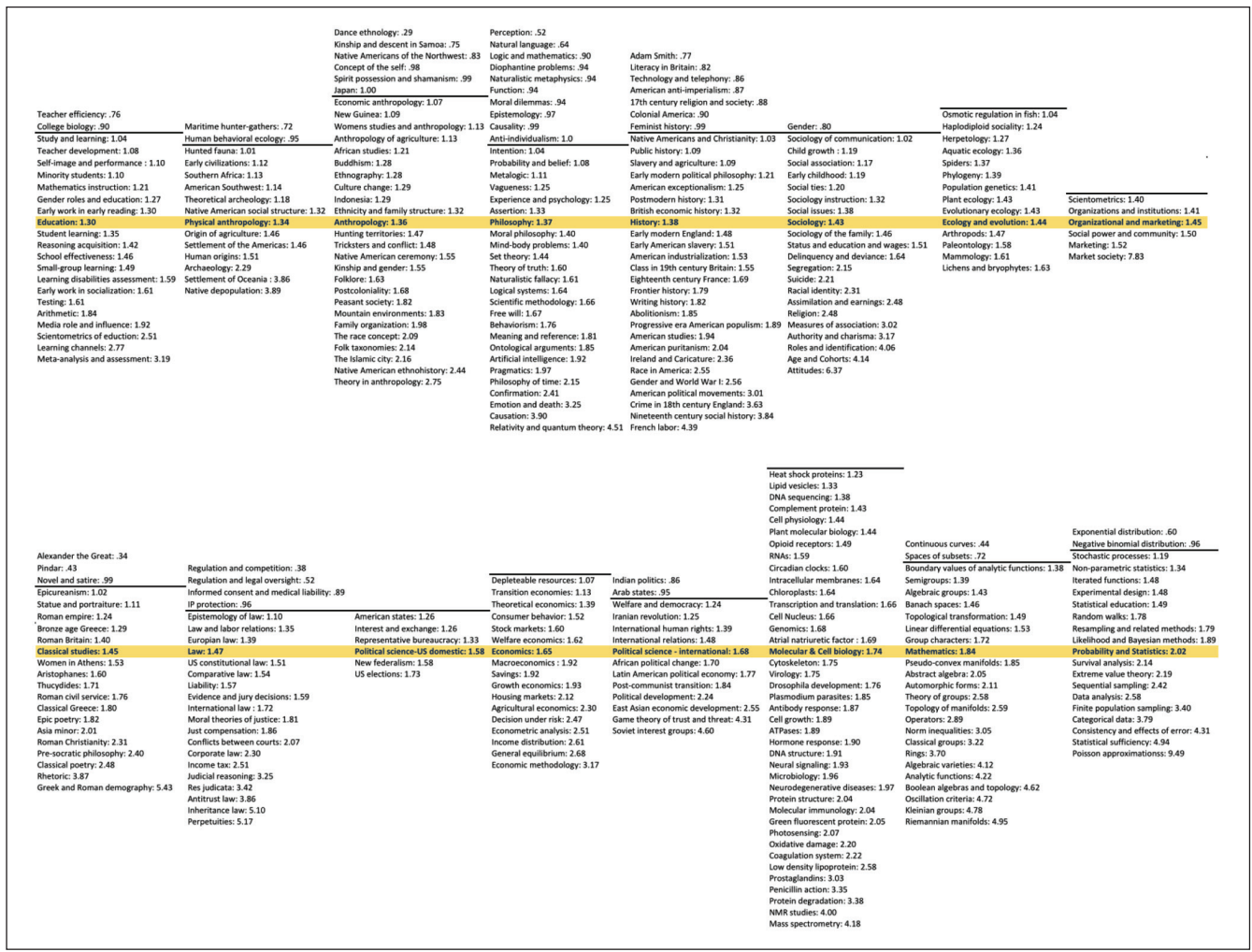

**Figure 7.** Ratios of men's to women's self-citation rates by field and subfield, in JSTOR articles, 1779–2011. Yellow center line represents self-citation ratio for overall field, with women's self-citation rate set at 1.0. Each subfield is arrayed around its corresponding field on the basis of the subfield's ratio compared with the larger field's ratio. The solid line within each column represents the location of an equal ratio (1.0) of self-citations between men authors and women authors.

Figure 7 shows the relative self-citation ratios at the field level. For each of these 16 largest fields, we also display the ratios for the subfields determined by the hierarchical map equation algorithm. Even within each major academic research field, gender ratios of self-citation vary depending on the subfield. Some subfields fall above the line indicating a ratio of 1.0, indicating that women self-cite more on average than men in that subfield. So that readers can explore these results for themselves, we present self-citation rates by gender across research domains in an interactive data visualization at http://eigenfactor.org/gender/self-citation/.[16]

---

[16]The browser allows the viewer to click to zoom in to any field and view the rates of self-citation by gender across major fields and subfields.

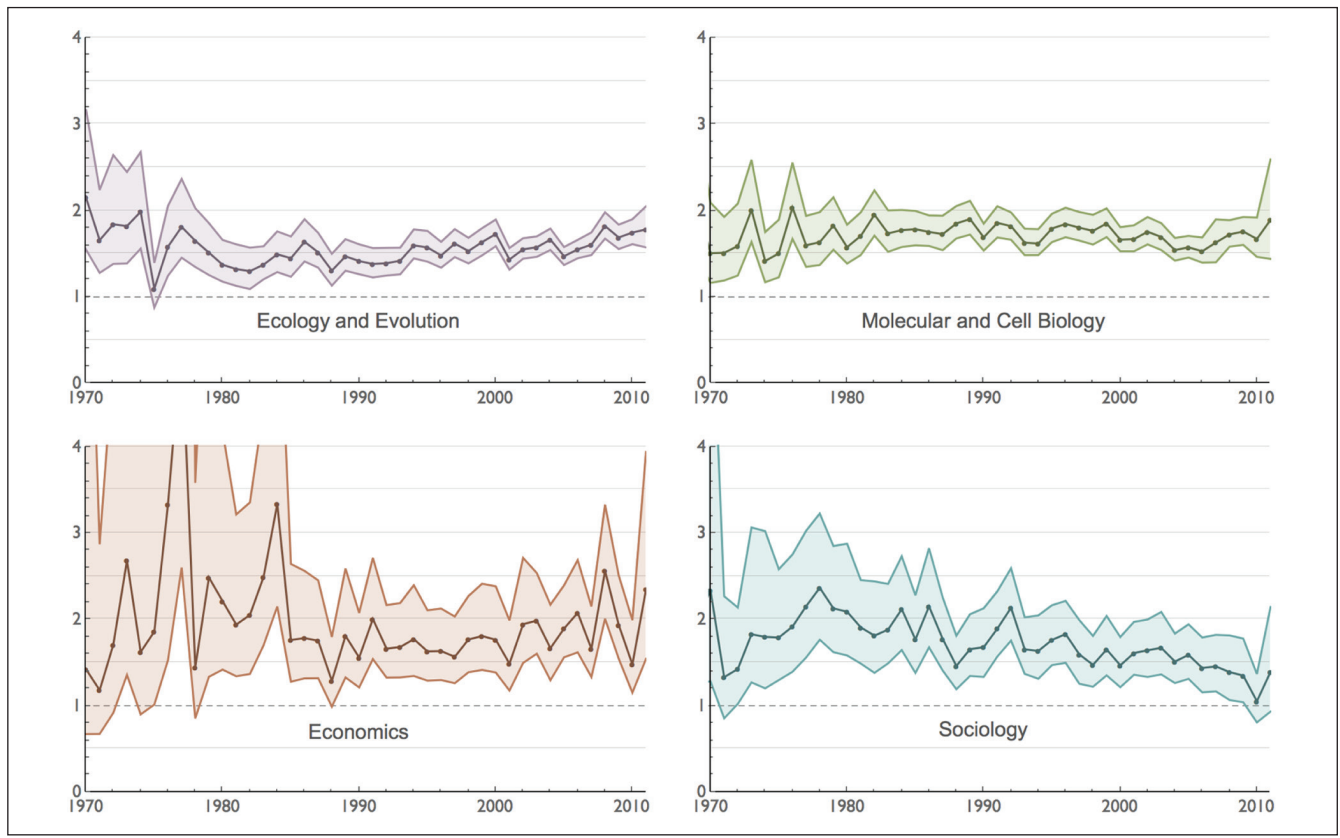

**Figure 8.** Gender ratios of self-citation rates across time for the four largest fields. The ratio of men's self-citations per authorship relative to women's self-citations per authorship, in the four largest JSTOR fields of ecology and evolution, molecular and cell biology, economics, and sociology over the period from 1970 to 2011. If men and women cited themselves at equal rates, the ratio shown would be 1.0. A value of 1.5 means that men in that field cite themselves 50 percent more than women in papers published during that year. Shaded intervals represent 95 percent bootstrap confidence limits.

### Self-citation Rates by Field over Time

We also looked at the changes in the self-citation rates and ratios within fields across time. This analysis helps ensure that our results are not an artifact of the different norms for self-citation in different disciplines combining with different proportions of men in each discipline.

If we look over time, we again see a consistent gender gap within each field. Figure 8 illustrates self-citation ratios across time for ecology and evolution, molecular and cell biology, economics, and sociology. Because the sample sizes for individual fields are substantially smaller than those for the entire corpus and because confidence intervals for ratios are sensitive to small sample sizes, we restrict our visualization to the most recent 40 years, for which we have the most data. For these four largest fields for which we have the best longitudinal data in the JSTOR data set, gender inequality in the self-citation ratio persists across the 40 years shown.

We also break down the rates of self-citation for men and women by the top 16 largest fields (Figure 9). The confidence intervals for the individual rates are naturally tighter than those for their ratios. Here we see that men's self-citation rate is generally higher than women's self-citation rate across time. In the fields (such as mathematics) and time periods (prior to 1970) with fewer papers, the confidence intervals do overlap more.

### Self-citation Rates by Size of Author Team

We wondered whether the tendency of men and women to collaborate and coauthor at different rates (Abramo, D'Angelo, and Murgia 2013; Bozeman and Gaughan 2011; Zeng et al. 2016) and the lower likelihood of women to write sole-authored papers (West, Jacquet, et al. 2013) might play any role in the gender differences in self-citation rates. To explore this question, we looked at the differences in the mean number of self-citations per authorship across papers with 1 to 20 authors. Figure 10 illustrates that those with sole-authored papers and with smaller teams of collaborators have a higher mean number of self-citations. Author-to-author self-citations occur at lower rates in papers with more authors. However, there were no interactions with gender.

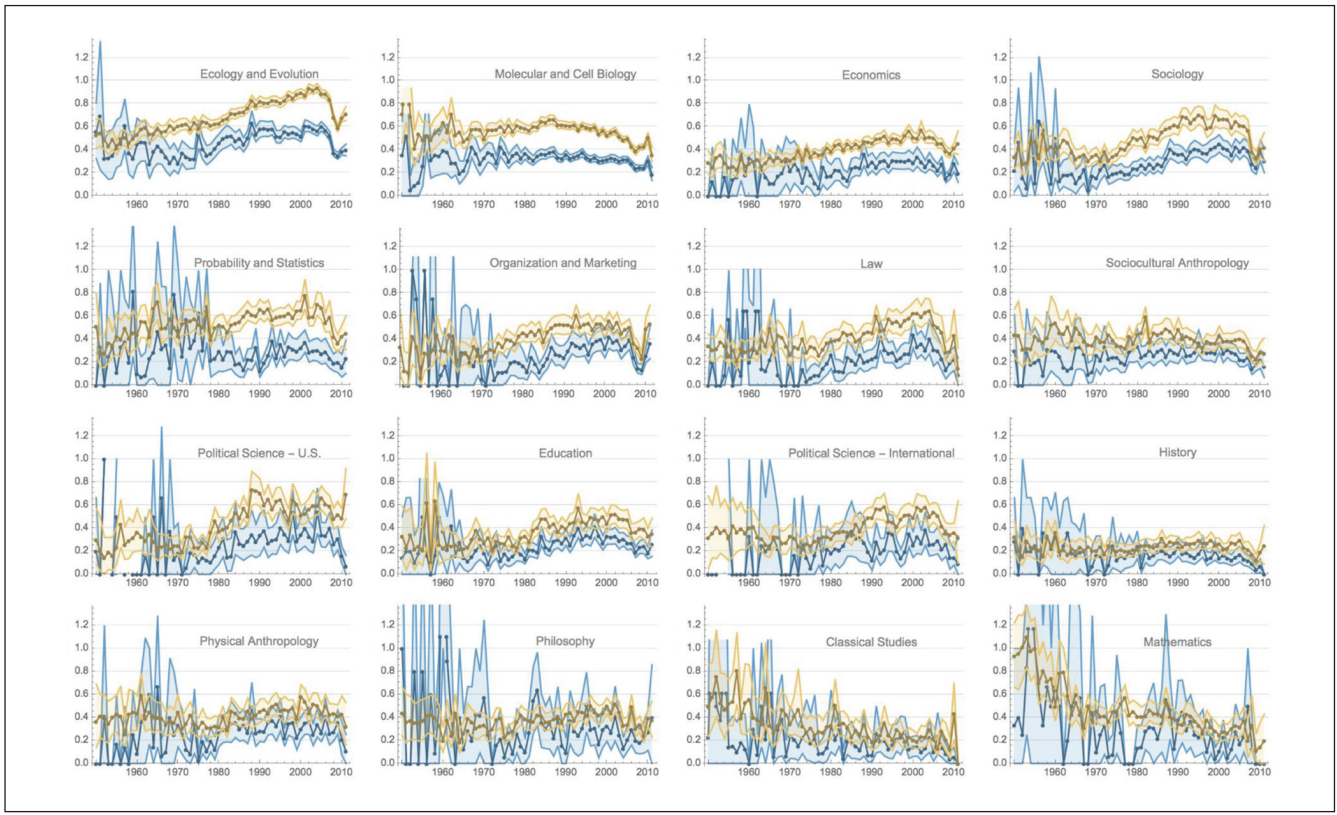

**Figure 9.** Men consistently self-cite more than women across fields. Shown here are the mean number of men's self-citations per authorship (yellow line) and women's self-citations per authorship (blue line) for the 16 largest fields in the JSTOR data set over the period from 1970 to 2011. Shaded intervals represent 95 percent bootstrap confidence limits.

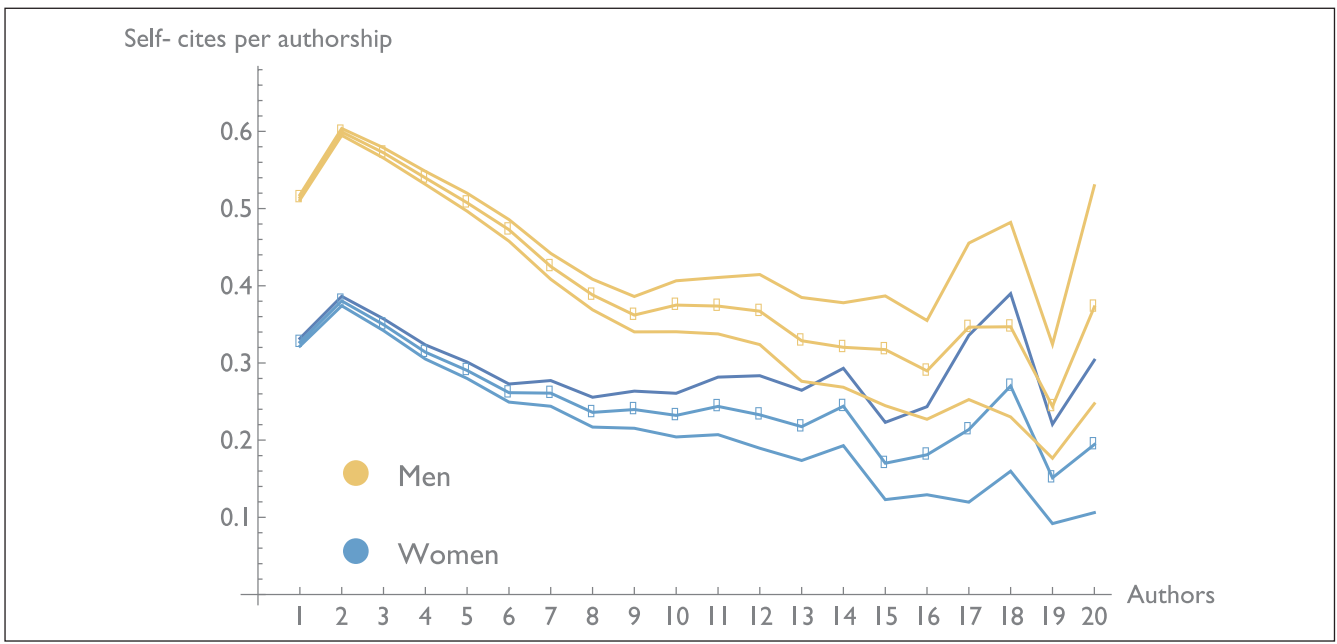

**Figure 10.** Mean number of self-citations per authorship by the number of authors on a paper, in JSTOR, 1779–2011. We truncate the results at 20 authors because, given small sample sizes, the data become excessively noisy beyond that threshold. A value of 1.0 on the vertical axis indicates that, on average, each author cites one of his or her previous papers in the current paper. Shaded intervals represent 95 percent bootstrap confidence limits.

## Discussion

Our study uses an unprecedentedly large data set of 1.5 million papers across a broad range of academic fields to examine trends in self-citation by academic researchers. Examining 39.4 million author-to-author citations and more than 1 million self-citations in the JSTOR database, we uncovered a number of important patterns:

1. About 9.4 percent of all citations are self-citations that reference previous papers written by one of or more of the current paper's authors. This indicates that self-citations have the potential to make up an important fraction of all citations to authors' work.
2. Compared with women, men are more than 10 percentage points more likely to self-cite (21.2 percent of women authorships vs. 31.4 percent of men authorships self-cite). Still, the majority of authors never cite themselves in a given paper.
3. In the last two decades of our data, men cited themselves at 1.7 times the rate of women.
4. There is wide variation across fields and subfields, but we do not observe any obvious relationship between the proportion of women in a field and the relative rates of women's and men's self-citation in that field.

### *Potential Mechanisms*

Why might men academics cite their own previous work more than women academics? Although our JSTOR data include a large number of papers and self-citations, they do not contain variables that allow us to determine the cause of the patterns we identify. However, prior research suggests several mechanisms that are consistent with our results. We review five mechanisms here, which potentially contribute to the gender self-citation gap and therefore, ultimately, to the cumulative disadvantage that women face in achieving equal recognition for their knowledge in the academic workplace:

1. Men may self-cite more because they evaluate their abilities more positively than women.
2. Men face fewer social penalties for self-promotion.
3. Men specialize more in academic subfields, and specialization may encourage more self-citation.
4. Men publish more papers, particularly earlier in their careers, and therefore have more work to cite.
5. Men publish different types of papers, namely, the types of papers an academic may be more likely to self-cite.

These mechanisms are by no means mutually exclusive, and we consider it likely that several may contribute to the gender gap we have observed. Some potential explanations are more behavioral, others more structural; our data do not allow us to conclusively adjudicate among them. We describe the existing evidence for each in turn.

The first two mechanisms—women's lower self-assessments of their accomplishments and greater social sanctions against women who self-promote—are related. Status beliefs about gender shape men's and women's behavior and expectations of themselves and others (Ridgeway 2001, 2014). Because women are perceived as lower status, they are often evaluated more negatively than equally qualified men candidates, by women as well as by men (Moss-Racusin et al. 2012; Reuben, Sapienza, and Zingales 2014). Women evaluate their own abilities more critically, even when faced with evidence of equivalent performance (Correll 2001, 2004). Women are especially prone to be evaluated critically (Cech et al. 2011; Thébaud 2010) or penalized for success (Heilman et al. 2004) when working in male-dominated domains. However, recall that we did not find that women self-cited less in more male-dominated fields. We found no relationship between the proportion of men in a field and the likelihood that a woman will self-cite. However, academia overall is male-dominated. If social sanctions for self-promotion are playing a role in women's lower likelihood to self-cite, then, at least according to our results, they are likely exerted in a more generalized way; that is, women are being sanctioned within academia or society as a whole, rather than by field.

When women seek to actively establish their competence by self-promoting, they often experience backlash from both men and women (Rudman et al. 2012). Gendered perceptions of self-promotion likely influence perceptions of self-citation, which could be viewed as a form of self-promotion in the academic workplace. Women are less likely than men to negotiate for what they want in the workplace. Men are also more likely to receive the corresponding rewards from these negotiations, such as higher salaries (Babcock and Laschever 2007; Babcock et al. 2003). Status expectations are particularly likely to operate in ambiguous contexts where evaluation criteria are subjective and loosely defined (Fox 2001; Ridgeway 2011)—such as those surrounding evaluations of the importance of an academic paper.

Field segregation by gender may also contribute to gender discrepancies in self-citation rates, for two reasons. First, fields have different norms around self-citation. Self-citation rates are higher in the natural sciences (Snyder and Bonzi 1998). We might expect to find higher self-citation rates in fields with more men authors. However, this is not the case: comparing across fields, there is no significant correlation between the mean number of self-citations per paper and the fraction of men authors in a field. Second, men tend to specialize more within their academic fields, at least within the disciplines of sociology and linguistics (Leahey 2006); this more specific focus may encourage self-citation. A research strategy whereby a scholar is focusing on building on previous work would likely result in many more self-citations. One remaining question for future research is whether specialization might explain gender differences in self-citation tendencies; we hope to test this in future work.

In part because men specialize more (Leahey 2006, 2007), several studies found that women faculty members tend to publish fewer papers than men faculty members (Barnett et al. 1998; Bentley and Adamson 2004; Cole and Singer 1991; Fox 2005; Symonds et al. 2006). Not only does higher productivity lead to more papers for scholars to self-cite; more productive scholars also generate more highly cited papers (Symonds et al. 2006). Gender differences in publication productivity vary depending on measures used, field, controls included, and time period studied (Bentley and Adamson 2004; Kyvik and Teigen 1996; Weisshaar forthcoming; Xie and Shauman 1998). Others find that this gender gap shrinks over the career trajectory (Long 1992) and that it has largely disappeared or reversed in more recent cohorts, with women publishing more than men (van Arensbergen et al. 2012; Xie and Shauman 1998).

As discussed earlier, the ratio of women to men in academic careers decreases as we climb up the academic status ladder. Attrition out of the academic pipeline means that women have fewer papers to self-cite and fewer later opportunities to do so, in aggregate in our data set.[17] This is one source of the productivity discrepancy because men will have overall greater productivity throughout their longer careers (in aggregate). Differences in this aggregate productivity might cause or further exacerbate gender inequality in self-citation counts,[18] resulting in cumulative disadvantage in apparent impact. Despite the increase in the number of women in more senior academic positions in many fields, we do not see a trend toward equality in women's and men's self-citation rates over time. We would expect this demographic shift to result in more papers for these senior women to self-cite. But our observations do not indicate any decrease in the self-citation gap over the past 50 years.

Finally, there are also differences in the types of papers produced by men and women; for instance, women are significantly underrepresented as authors of single-authored papers and, on papers with three or more authors, in the prestigious positions of first and last author (West, Jacquet, et al. 2013). These types of papers may constitute the kind of work that would be in the authors' core areas of research interest and thus papers they may be more likely to self-cite. The mean number of self-cites per authorship is also smaller for larger groups of coauthors (Figure 10). Because women are not publishing single-authored papers as often as men (West, Jacquet, et al. 2013), they are likely to have fewer self-citations per authorship. Overall, however, it may be that those types of papers that women tend to publish disproportionately fewer of are also those that attract more self-citations.

To provide another test of our findings, we look at self-citation by gender from the smaller Social Science Research Network (SSRN) data set. We find a self-citation gap of similar magnitude in this alternative data set, although there is no evidence that men with equal numbers of papers self-cite more than women. Methods and discussion of this analysis are available in Appendix A. The SSRN data set differs from the JSTOR data set on a number of important features: it is smaller, represents fewer disciplines, and is a non-peer-reviewed prepublication archive that only some authors in relevant fields elect to use, so there are many reasons to believe that selection into the database might affect results in key ways that are outside the scope of this study to explore.

## Implications

Whether the self-citation gap is ultimately a cause or a consequence—or some combination thereof—of the very real gender imbalances within academia is not something we can determine. Nor is such a determination necessary for the gender gap in self-citations to be an important descriptive finding. If the gender gap in self-citations is a consequence (or a cause) of a productivity gap, then we should be looking to understand the conditions under which this gap exists. The cumulative disadvantage that results from the self-citation gap is yet another way that a productivity deficit would harm women academics. On the other extreme of the structural-behavioral spectrum, we might find that there is no productivity difference and the self-citation gap is due entirely to gender differences in self-citation behavior. Then, more appropriate remedies might be adjusting for self-citations in hiring and promotion, for example. Either way, the large and ongoing gender gap in self-citations is an important aspect of gender parity in academic careers, especially because the disparities caused by self-citation gaps worsen over time.

---

[17]For example, in our data set, a higher proportion of the women represented may have gone on to nonacademic careers than the men in our data set. If this were the case, those women who published as PhD students or as young researchers would not have had as much opportunity to self-cite later in their careers as the men who stayed on to become career academics.

[18]Here is a highly simplified example of how we could get the results above without any difference in self-citation behavior. Suppose men and women behave the same, such that there is no gender-differentiated effect of self-promoting behavior (mechanism 1) or social sanctions for self-promotion (mechanism 2). Imagine that everyone cites everything they have ever written in every paper they write. But suppose the distribution of paper counts differs. All women only ever write two papers, while all men only ever write three. In this example, the average number of self-citations per authorship for men is 1 (each man cites his first paper in his second paper, and his first two papers in his third paper, for a total of three self-citations across three papers). The average number of self-citations per authorship for women is 0.5 (each woman cites her first paper in her second paper, for a total of one self-citation across two papers). The gap in overall self-citation rates would diminish with increasing numbers of papers, but as long as men published even slightly more papers—on average—than women, and both self-cited at the same rates, there would always remain a difference in the average self-citation rates by gender.

Citation follows a pattern of preferential attachment: the tendency for new citations to refer to papers that are already well cited (Barabási et al. 2001; de Solla Price 1976). Thus, self-citation increases the number of citations from others (Fowler and Aksnes 2007) in a process of cumulative advantage. The gender difference in self-citation is therefore likely to be a driver of gender differences in numbers of citations received from other authors. And this has consequences beyond scholarly recognition. An academic's visibility, reflected in citation counts, has a significant, direct, positive effect on his or her salary. In a study of linguists and sociologists, visibility explained half of the $13,000 salary gap between men and women (Leahey 2007).

The motives for self-citation vary (Hyland 2003; Safer and Tang 2009; Tang and Safer 2008), but self-citation is one of the few direct ways an academic can increase his or her own citation count. Some scholarly databases (e.g. the Thomson Reuters Web of Science) provide a separate count of self-citations, while others (e.g., Google Scholar) do not. However, merely encouraging women to cite their own work more is not a simple solution: it may have unintended consequences due to backlash against women's self-promotion (Rudman 1998). Furthermore, insisting that scholars self-cite more in order to enhance their reputations could increase irrelevant self-citations. Should this happen, it would become even more difficult to make accurate judgments of the quality and influence of a scholarly work.

When interpreting the impact metrics of scholars' work, university hiring and tenure committees should be aware that women are likely to cite their own work less often. Considering other proposed measures for scientific impact that exclude self-citation (e.g., Ferrara and Romero 2013; West, Jensen, et al. 2013)—and adjust for the cumulative advantage of additional citations that accrue as a result of those self-citations—could make evaluation processes less gender-biased and improve equity in the academic community.

## Appendix A

### The SSRN Data Set: Additional Verification of the Gender Gap

To provide another test of our findings, we look at self-citation by gender from the smaller SSRN data set. This set of papers is unusual in that the authors have been carefully disambiguated (see West, Jensen, et al. 2013): we can distinguish for example between one individual named Rita Martin who has written two papers and two individuals named John Williams, each of whom has written one. The SSRN data set includes 426,412 papers (including preprints) from 99,465 authors, with more than 2.4 million citations among those papers. In addition to being smaller, the SSRN data set differs from JSTOR because authors voluntarily upload papers to SSRN.

We follow the same procedure for gender assignment as with the JSTOR data. Men account for 73 percent (38,265) of authors who can be disambiguated by name and whose gender can be identified, while women account for 27 percent (14,379). We can identify a gender for a total of 10,212,014 authorship-to-authorship citations. Men authors have 280,818 papers with 181,742 self-citations, for an average of 0.647 self-citations per paper. Women authors have 68,256 papers with 28,075 self-citations, for an average of 0.411 self-citations per paper. Among all authors, including those with zero self-citations, men self-cite an average 0.193 and women 0.128 times per paper.

In our preliminary evaluation of the SSRN database, we found a gender self-citation gap of equivalent magnitude to the one we find in JSTOR. In the SSRN data, men make up 73 percent of authorships but 87 percent of self-citations. However, the SSRN data do not support the hypotheses that this gap arises because men and women behave differently in terms of self-citation. Men with *n* papers in the SSRN database do not appear to self-cite appreciably more than women with *n* papers. However, the self-citation gap in the SSRN data set could arise because men authors have more citation targets or because men and women who voluntarily submit papers to the SSRN are not representative of academics, more generally. The SSRN data set differs from the JSTOR data set on a number of important features: it is smaller, represents fewer disciplines, and is a non-peer-reviewed prepublication archive that only some authors in relevant fields elect to use, so there are many reasons to believe that selection into the database might affect results in key ways that are outside the scope of this study to explore. However, the finding of a similar self-citation gap in a very different data set is nonetheless reassuring to our results.

## Appendix B

### Continuous Levels versus Threshold Effects of Proportion Women on Self-citation

Kanter (1993) theorized that the threshold effect would appear when the proportion of women in a field reached 15 percent. (The 15 percent threshold Kanter suggested has not really held up empirically over time, but it seems a reasonable point to start. Others have suggested 20 percent, still others 30 percent.) In our data, there is no cutoff at which there appears a threshold effect. The cutoff of 15 percent women authorships in a field lies between law and economics (in law, 15.2 percent of authorships are women). Because we cannot disambiguate our authors, this analysis is done in authorships, not proportion of the field that is women. We suspect that for the proportion of women at work, there is a threshold that varies with context. Figures B1 and B2 are ordered with decreasing proportion of women authorships, from most at the top to least at the bottom.

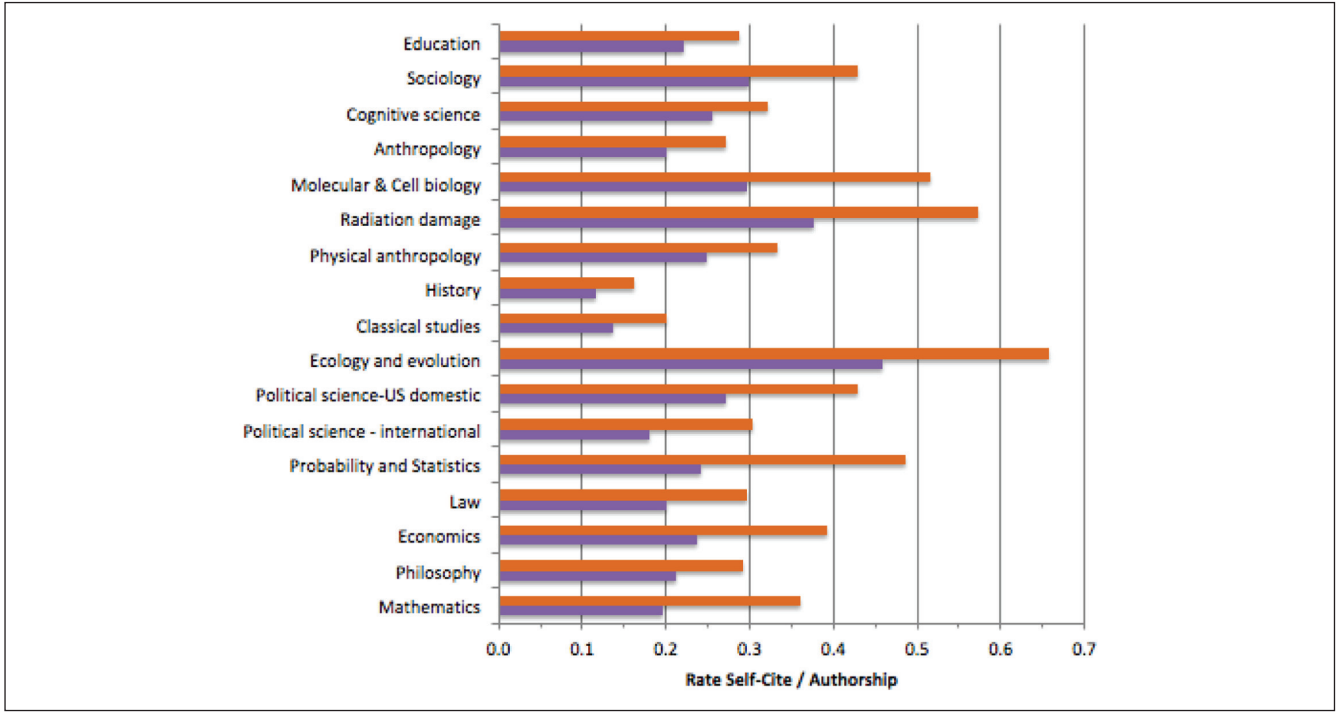

**Figure B1.** Rate of self-citation per authorship across fields. Fields are ordered from highest proportion of women authorships at the top (education) to the lowest proportion of women authorships at the bottom (mathematics), as a proxy for the proportion of women in the field overall.

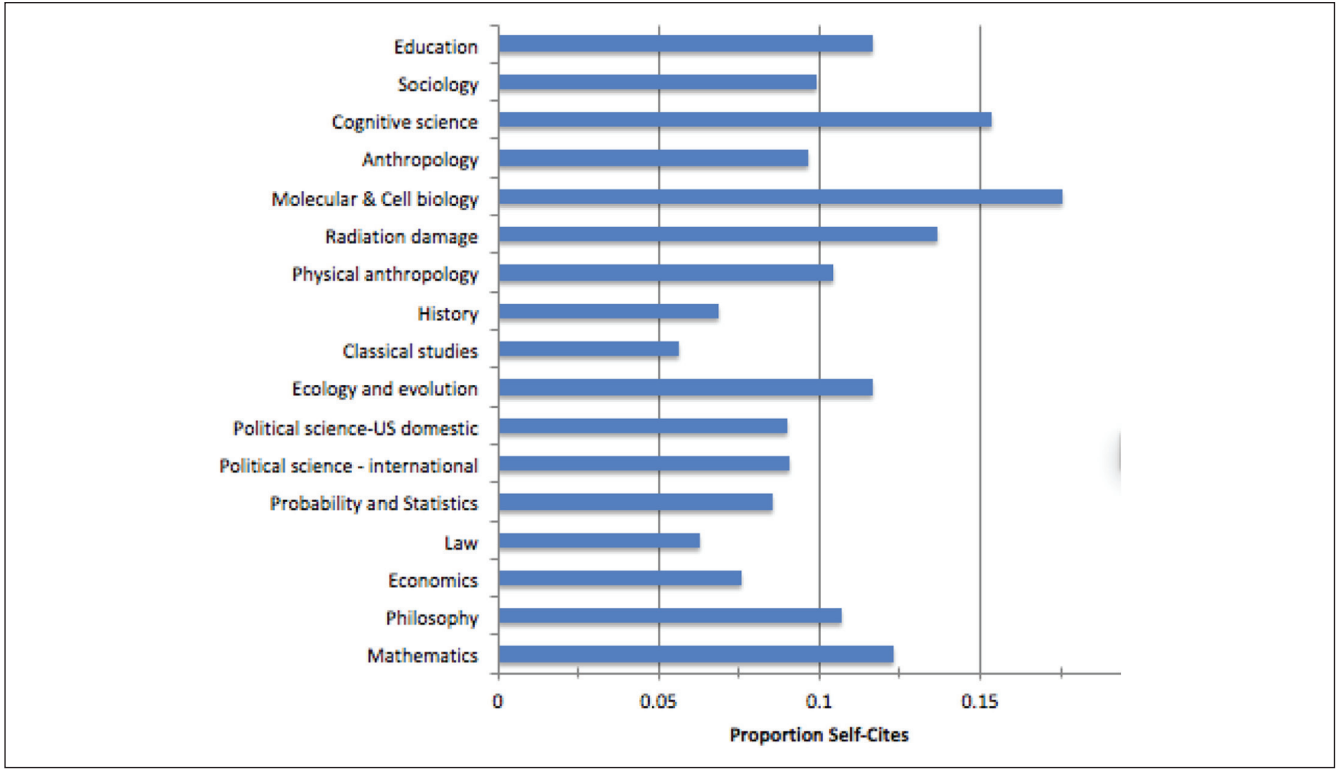

**Figure B2.** Proportion of self-citations across fields. Fields are ordered from highest proportion of women authorships at the top (education) to the lowest proportion of women authorships at the bottom (mathematics), as a proxy for the proportion of women in the field overall.

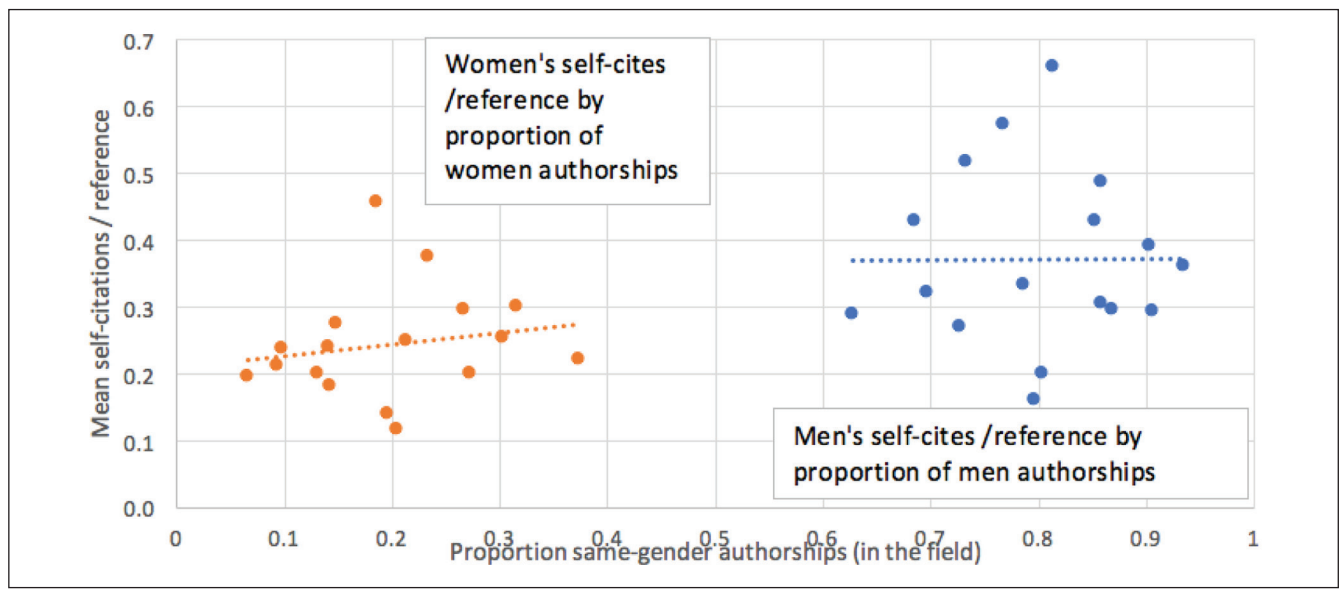

**Figure B3.** Rate of self-citations per reference by proportion of same-gender authorships in field. The rate of self-citations is normalized by the number of references per authorship, to account for differential citation practices between fields.

Figure B3 shows the rate of self-citation, separately for men and women authorships, by the proportion of men and women authorships in each field of study (for top fields). We do see a slight upward linear effect of the rate of women's self-citation correlated with the proportion of women in a field, but with so few data points, it is impossible to call this a clear effect.

### Acknowledgments

We thank JSTOR for providing access to the data used here and for compiling the citation graph used to assign papers to fields. We thank Erin Leahey, Erin Cech, David Miller, Flo Débarre, and Howard Aldrich for feedback on earlier drafts of this article. We also thank the Mathematica Stack Exchange community for assistance in creating Figure 3 and Pooja Loftus, Armand Rundquist, and Jason Hirshman for technical and statistical assistance.

### Funding

The author(s) disclosed receipt of the following financial support for the research, authorship, and/or publication of this article: This work was supported in part by a National Science Foundation Graduate Research Fellowship (grant DGE-1147470) to M.M.K. and a John Templeton Foundation Metaknowledge Network grant to J.D.W. and C.T.B.

### Supplemental Data

The raw data for the study have been provided to us by JSTOR (http://www.jstor.org) under a data use license with the University of Washington. We are committed to making the data as open as allowed under the terms of this data use agreement. To put our use of the data in context, we use the JSTOR data for two purposes: (1) to hierarchically classify the roughly 2 million papers in our data set into fields, subfields, and so on, and (2) to determine gender patterns of authorship in these fields. The hierarchical categorization of the fields and subfields of academia is the same as that in our previous *PLoS ONE* paper, "The Role of Gender in Scholarly Authorship" (http://journals.plos.org/plosone/article/file?id=10.1371/journal.pone.0066212&type=printable). With JSTOR's blessing, we have developed an in-depth Web-based data explorer to accompany the present paper, available at http://www.eigenfactor.org/gender/self-citation/. This reveals both the full structure of the scientific fields and subfields and the associated gender ratios for each.

## Author Biographies

**Molly M. King** is a PhD candidate in sociology at Stanford University and a National Science Foundation Graduate Research Fellow. Her research investigates who "owns" knowledge and the implications of that for inequality. Her dissertation looks at what people know and how that is affected by class, gender and race. She also coauthored "The Role of Gender in Scholarly Authorship" (in *PLoS ONE*). Her research agenda looks at how knowledge is socially structured.

**Carl T. Bergstrom** is a professor of biology at the University of Washington. He received his PhD in biological sciences from Stanford University. His research focuses on information in biological and social systems: how do these systems evolve or otherwise develop ways of acquiring, storing, processing, and transmitting information? His recent projects include contributions to the game theory of communication, applications of information theory to understanding complex networks, studies of scientific communication, and more applied work in the area of disease evolution. He is the coauthor (with Lee Dugatkin) of a leading undergraduate textbook, *Evolution*.

**Shelley J. Correll** is a professor of sociology and, by courtesy, organizational behavior at Stanford University, where she also serves as the Barbara D. Finberg Director of the Clayman Institute for Gender Research. Her current research examines how legal mandates and organizational policies and practices can reduce the effect of status beliefs on workplace gender inequalities and how gender status beliefs produce inequalities in product markets.

**Jennifer Jacquet** is an assistant professor and social scientist in the Department of Environmental Studies at New York University. She works on large-scale, transboundary cooperation dilemmas, including overfishing, climate change, and the illegal wildlife trade, and is also interested in questions related to gender in academia. She is the author of *Is Shame Necessary? New Uses for an Old Tool*.

**Jevin D. West** is an assistant professor at the Information School at the University of Washington, codirector of the DataLab, and a Data Science Fellow at the eScience Institute. He received his PhD in biology at the University of Washington and continued with a postdoctoral fellowship at the Department of Physics at Umea University in Sweden, where he worked on the mapequation.org project in the Icelab. He cofounded Eigenfactor.org, a research project that maps large-scale citation networks in order to better understand, navigate, and evaluate the scientific literature. His research in the science of science includes the evolution of scientific disciplines, economics of scholarly publishing, data mining of large corpora, and the sociology of science.